\documentclass[aps,letterpaper,twocolumn,prl,footinbib,floatfix,showpacs]{revtex4}
\usepackage{amsmath}
\usepackage{amsfonts}
\usepackage{amssymb}
\usepackage[scanall]{psfrag}
\usepackage{psfrag}
\usepackage{graphicx}

\begin{document}
\newcommand{\of}[1]{\left( #1 \right)}
\newcommand{\sqof}[1]{\left[ #1 \right]}
\newcommand{\abs}[1]{\left| #1 \right|}
\newcommand{\avg}[1]{\left< #1 \right>}
\newcommand{\cuof}[1]{\left \{ #1 \right \} }
\newcommand{\pil}{\frac{\pi}{L}}
\newcommand{\bx}{\mathbf{x}}
\newcommand{\by}{\mathbf{y}}
\newcommand{\bk}{\mathbf{k}}
\newcommand{\bp}{\mathbf{p}}
\newcommand{\bl}{\mathbf{l}}
\newcommand{\bq}{\mathbf{q}}
\author{Eliot Kapit and Erich Mueller}
\affiliation{Laboratory of Atomic and Solid State Physics, Cornell University}
\title{Exact Parent Hamiltonian for the Quantum Hall States in a Lattice}
\begin{abstract}

We study lattice models of charged particles in uniform magnetic fields. We show how longer range hopping can be engineered to produce 
 a massively degenerate manifold of single-particle ground states with wavefunctions identical to those making up the 
  lowest Landau level of continuum electrons in a magnetic field. We find that in the presence of local interactions, and at the appropriate filling factors, Laughlin's fractional quantum Hall wavefunction is an exact many-body ground state of our lattice model.  The hopping matrix elements in our model fall off as a Gaussian, and when the flux per plaquette is small compared to the fundamental flux quantum one only needs to include nearest and next nearest neighbor hoppings.  We suggest how to realize this model using atoms in optical lattices, and describe observable consequences of the resulting fractional quantum Hall physics.
\end{abstract}

\pacs{03.75.Lm,67.85.Hj,03.75.Hh,73.43.-f}
\date{\today}

\maketitle


The interplay between periodic potentials and magnetic fields is an important topic \cite{hofstadter,sorensen,palmerkleinjaksch,moller,cooper}.  In the tight binding limit, the lattice broadens the Landau levels into a series of finite bandwidth ``Hofstadter bands" which can be represented as a self-similar fractal.  Since the original band-gaps persist,  the integer quantum Hall effects are robust against the lattice.  The split degeneracy, however, invalidates  many of the analytic arguments used to explain the fractional quantum Hall effect \cite{laughlinoriginal,haldaneexact,girvin,yoshoika}, and  questions remain about the nature of the interacting system.  Here, by adding longer range hoppings to a Hubbard model, we produce a Hamiltonian for which several Hofstadter bands coalesce into a single degenerate manifold.  Adding local repulsion between the particles, we show that at appropriate filling factors the Laughlin wavefunction becomes an exact ground-state.

In a uniform magnetic field, 
the most general hopping Hamiltonian on a two-dimensional square lattice is
\begin{align}\label{H}
H = \sum_{j \neq k} J& \of{z_{j},z_{k}} a_{j}^{\dagger} a_{k}, \\
J \of{z_{j},z_k} &= W(z) e^{ (\pi/2) \of{z_{j} z^{*} - z_{j}^{*} z }  \phi },\nonumber
\end{align}
where the position of the $j$'th lattice site is written in complex notation as $z_j=x_j+i y_j$, and $z=z_k-z_j$.  The operators $a_j$ annihilate an atom at site $j$.
The  phase factor 
 $\of{z_{j} z^{*} - z_{j}^{*} z }\phi  = 2i \of{x_{j} y - y_{j} x}\phi $, corresponds to a uniform magnetic field in the symmetric gauge, with flux $\phi$ through each plaquette.  This flux is only defined modulo 1, and having a full flux quantum through each plaquette is gauge equivalent to no flux. We will explicitly assume $0 \leq \phi \leq 1$, and take $\phi = p/q$ to be the ratio of two relatively prime integers.
If one chooses $W$ to be $-t$ for nearest neighbors and zero otherwise, one reproduces the Hofstadter spectrum \cite{hofstadter}.
We show that if instead we choose
\begin{align}\label{W}
W(z)&= t\times G \of{z} e^{ -\frac{\pi}{2} \sqof{ \of{1-\phi} \abs{z}^{2}}}\\
G \of{z} &\equiv \of{-1}^{x+y+x y},\nonumber
\end{align}
the lowest $p$ Hofstadter bands collapses to a single fully degenerate Landau level.
 Although we work in the symmetric gauge $\mathbf{A} = \of{ B/2} \of{x \hat{\bf y} - y \hat{\bf x}}$,  converting our results to other gauges is trivial: under a gauge transformation $\mathbf{A(r)}\to \mathbf{A({\bf r})}+\nabla\Lambda({\bf r})$ and $c_j\to c_j e^{i\Lambda({\bf r_j})}$.
 The flux is measured in units of $\phi_0=h/e$, where $h$ is Planck's constant, and $e$ is the electric charge.  Our derivation of this Hamiltonian is similar to one used by Laughlin \cite{LaughlinSum} and subsequently corrected/extended in \cite{schroeter1,schroeter2}. The paradigm of creating a parent Hamiltonian for which a desired quantum state is an exact eigenstate has been fruitful in a number of other spin models \cite{auerbach,affleck,kivelson}. We work in units where $t=1$. A similar construction can be defined for triangular lattices \cite{trianglefoot}.

Our results have deep implications.  First and foremost, they provide an exact equivalence between the lowest Landau level in the continuum and in a realistic lattice system.  This equivalence is unexpected, and can be further exploited.  For example, it provides an avenue for robust lattice calculations of continuum quantum Hall systems, and investigations of lattice fractional quantum Hall states \cite{moller,greiterthomale}.  Similarly it provides a means for lattice experiments to emulate an important continuum problem \cite{bloch}.

Massive degeneracies, such as the ones we found here, are related to symmetries.  Further theoretical work on this model may reveal these symmetries.  Our model also provides a convenient inroad towards a semiclassical understanding of the Hofstadter spectrum, and connections to magnetic breakdown in real materials \cite{gvozdikov}.

\begin{figure}
\includegraphics[width=\columnwidth]{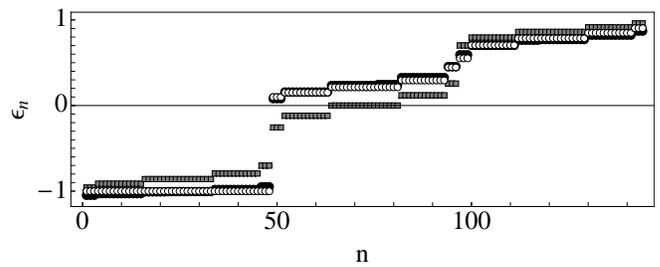}
\caption{All single particle eigenvalues for the hopping Hamiltonian in Eq.~(\ref{H}) with $\phi = 1/3$ on a $12\times12$ lattice with periodic boundary conditions. The index $n$ labels eigenvalues from smallest to largest.  The white disks  use hopping matrix elements given by Eq.~(\ref{W}),
 the black disks are the same model with only nearest and next nearest neighbor hopping, and the  grey boxes have only nearest neighbor hopping (the Hofstadter Hamiltonian).  Energies are all measured in units of $t$. The energies of the white and black disks are nearly indistinguishable, and the white disks obscure some of the black ones.  The lowest $1/3$ of the white disks are all degenerate.
 }\label{eigenstates}
\end{figure}

The most promising experimental realization of our model is in optical lattices \cite{bloch}. Optical lattice experiments can study both bosonic and fermionic quantum Hall states, and allow us in principle to study much larger fluxes than can be achieved with real magnetic fields. The gauge potential in the optical lattice system can be created in a number of ways: time-varying hopping elements \cite{sorensen}, lattices with multiple sets of minima \cite{mueller2004}, coherent Raman scattering \cite{spielman} and rotation \cite{hafezi,tung,williamsrot,williams,cooper,lin}. Further, optical lattice systems allow us to directly tune the hopping amplitudes between nearby sites. Long range hopping is difficult to arrange, but in our model $J$ falls off as a Gaussian, and in the limit of small $\phi$ it suffices to include only nearest and next-nearest neighbor hopping. The ratio of these hopping matrix elements can be controlled in an experiment by adjusting the shapes of the barriers between those sites.   One practical scheme would involve adding an additional array of shallow wells displaced by half a lattice spacing in both the $x$ and $y$ directions.  Integrating out these shallow sites will renormalize the nearest and next-nearest neighbor hoppings.  A second scheme would be to divide the original lattice into two sublattices.  Separating the sublattices in the $z$-direction will attenuate the nearest neighbor tunelling while leaving the next-nearest neighbor matrix element largely unchanged. Figure~\ref{eigenstates} compares the energies of the single particle eigenstates of Eq.~(\ref{H}), using the $W$ in Eq.~(\ref{W}), as well as truncating to only nearest neighbors or next-nearest neighbors.  As one can see, even for $\phi=1/3$, the next-nearest neighbor hopping already reduces the bandwidth to $0.1 t$.

Another possible realization of Eq.~(\ref{W}) would be 2D electron gases in a superlattice \cite{melinte,geisler,feil,hugger}. Hopping amplitudes can be tuned through altering the device structure. These systems also naturally include long-range Coulomb interactions, which are absent in trapped neutral atoms, and lead to richer many-body physics.


 Not only does this Hamiltonian produce a macroscopically degenerate manifold of single particle ground states, but this manifold is spanned by wavefunctions of the form
%
%
\begin{eqnarray}\label{wf}
\psi_{n} \of{z_j} =\langle j|\psi_n\rangle= z_j^{n} \exp \of{ - \frac{\pi \phi}{2} \abs{z_j}^{2}},
\end{eqnarray}
all with energy $\epsilon=-1$.  Remarkably, this is the same structure as the continuum problem, where the LLL is characterized by the same degenerate set of single particle states. To prove this result, we write
\begin{equation}\label{eigen}
\frac{ \left< j\right| H \left| \psi_{n} \right>   }{  \left< j | \psi_{n} \right>   }=\sum_{z \neq 0} G \of{z} \frac{ \of{z_{j} + z}^{n}  }{ z_{j}^{n}  } e^{-\frac{\pi}{2} \abs{z}^{2} - \pi \phi z_{j}^{*} z }.
\end{equation}
We then appeal to the singlet sum rule \cite{perelomov,LaughlinSum}, 
\begin{eqnarray}
k \of{c} \equiv \sum_{z}e^{c z} G \of{z} e^{- \frac{\pi}{2} \abs{z}^{2} } =  0  \quad \forall c,
\end{eqnarray}
where the sum is over all $z=n+i m$ with integer $n$ and $m$.
By taking any number of derivatives with respect to $c$ one finds
\begin{eqnarray}\label{sumrule}
\sum_{z} f \of{z} G \of{z} e^{- \frac{\pi}{2} \abs{z}^{2} } = 0,
\end{eqnarray}
for any \textit{entire} function $f \of{z}$ that diverges sufficiently slowly as $\abs{z} \rightarrow \infty$ 
\cite{entirefoot}.
Since we do not include the $z=0$ term in Eq. (\ref{eigen}), one immediately finds that the right hand side is simply -1, proving that the LLL wavefunctions (\ref{wf}) are degenerate eigenstates. No analogous argument works for the higher Landau level wavefunctions, which involve powers of both $z^*$ and $z$.

Given that the wavefunctions in (\ref{wf}) are identical to those of the continuum problem, the total number of degenerate states per unit area is the same as in the continuum; this results in $\phi N_s$ LLL wavefunctions in a region containing $N_s$ lattice sites.  Thus $\phi$ is the fraction of all single particle states which reside in the LLL.  Taking $\phi=p/q$, the standard Hofstadter problem yeilds $q$ distinct bands.  Thus, as we confirm numerically, our LLL must be made from the lowest $p$ of these.  This $p$-fold collapse is consistent with the relationship between the Chern numbers of the Hofstadter bands, and that of the LLL \cite{kohmoto}.

For $\phi>1/2$ it is natural to also consider the Hamiltonian formed if one replaces $\phi$ in Eq. (2) with $1-\phi$ and leaves equation (1) unchanged.  Due to the periodicity in $\phi$ of lattice models, this gives a Hamiltonian with the same absolute flux per plaquette, however it is clearly a distinct Hamiltonian, with shorter range hopping.  This alternative Hamiltonian yields states analogous to (\ref{wf}), but with $z$ replaced by $z^*$ and a degeneracy of $(1-\phi)N_{s}$.

The massive ground state degeneracy of our system can be lifted by interactions. Since our model reproduces the continuum lowest Landau level, we can simply use those results \cite{haldaneexact,yoshoika,halperin}. On-site repulsion in the lattice is equivalent to point interactions in the continuum. Consider for example the interacting Hamiltonian
\begin{eqnarray}\label{hint}
H = \sum_{j \neq k,\sigma} J \of{z_j, z_k} a_{j \sigma}^{\dagger} a_{k \sigma} + \sum_{j,\sigma,\sigma^\prime}  \frac{U_{\sigma\sigma^\prime}}{2}a_{j\sigma}^{\dagger} a_{j\sigma^\prime}^{\dagger} a_{j\sigma^\prime} a_{j\sigma}
\end{eqnarray}
where $U_{\sigma\sigma^\prime}>0$ is the on-site interaction energy of particles with spin states $\sigma$ and $\sigma^\prime$.
Any LLL wavefunction which vanishes when two particles coincide is a ground state of this Hamiltonian. Due to the structure of the LLL, there is a maximal atomic density for which this occurs. For single component bosons the highest density ground state is the $\nu=1/2$ Laughlin state, 
\begin{equation}\label{laughlin}
\psi(z_1,\cdots,z_N)=\prod_{i<j=1}^N (z_i-z_j)^2 \prod_{j=1}^Ne^{- \pi\phi/2 |z_j|^2}.
\end{equation}
For 2-component fermions it is the ferromagnetic ``111" state \cite{halperin}. At fixed density, these states are unique up to topological degeneracies. The other Laughlin states and more exotic quantum Hall states are also ground states if $\nu \leq 1/2$, however they are not unique if all interactions are local.

Longer ranged interactions (as found in electronic systems or in optical lattice experiments with dipolar gases \cite{hafezi})  will typically lift the degeneracy entirely \cite{assaad}. The subsequent analysis can be quite involved, but most continuum arguments will carry over.

Our construction is readily extended to a finite system with magneto-periodic boundary conditions,
\begin{equation}  
\psi \of{ z + n L + i m L}   = \psi \of{ z }  e^{i \pi\phi L \of{n y - m x} }.
\end{equation}
There one replaces the polynomials in Eq.~(\ref{wf}) with appropriate products of Gaussians and Jacobi theta functions \cite{haldane,AandS}.  
One also replaces $J(z_j,z)$ in Eq.~(\ref{H}) by its magneto-periodic extension
\begin{equation}\label{HP}
J_L(z_j,z)=\sum_R J(z_j,z+R)\exp\left(\frac{\pi}{2}(z_j R^*-z_j^*R)\phi \right),
\end{equation}
where the sum is over all $R=n L+ i m L$ for integer $n,m$.  This finite system is amenable to numerical calculations. To invoke the singlet sum rule in a periodic geometry, one must simply merge the sums on $z$ and $R$ into a single sum over all $z \neq 0$. The phase factors from the magnetoperiodicity of $\psi \of{z}$ and $J$ cancel each other.

\begin{figure}
\includegraphics[width=\columnwidth]{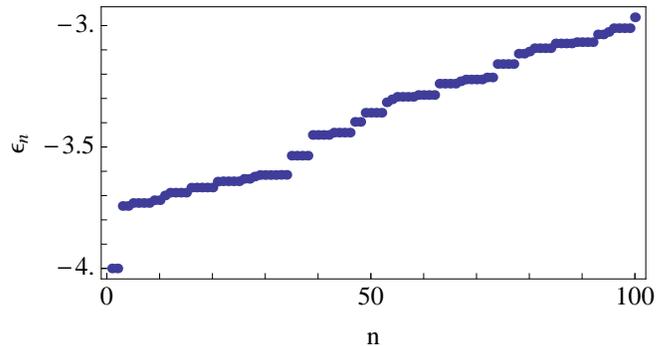}
\caption{First 100 eigenvalues for 4 particles on a $4\times4$ lattice with periodic boundary conditions, $\phi = 1/2$, and hard core repulsion. The two states at $\epsilon= -4$ are Laughlin states (\ref{wfn}); the degeneracy stems from the toroidal geometry. There is a distinct energy gap of $0.566 t_{nn}$ to the lowest excited states, where $t_{nn}$ is the nearest neighbor hopping amplitude.}\label{4x4}
\end{figure}

In the finite system with magneto-periodic boundary conditions
the $\nu=1/p$ Laughlin state is
\begin{eqnarray}\label{wfn}
\Psi \of{ \cuof{z_{n}} } &=&\Psi_{\rm cm}  \times\prod_{k < j}^{M} \chi_{jk}^p
\prod_{j=1}^M e^{ \pi \frac{p M}{2 L^{2}} \of{z_{j}^{2} - |z_{j}|^2}}\nonumber\\
\Psi_{\rm cm}&=& \prod_{i = 1}^{p} \theta_1\of{\pil\of{Z - Z_{i}}}.
 \end{eqnarray}
 The center of mass coordinate is $Z$ $=$ $\sum_j z_j$, and
$\chi_{jk}$ $=$ $\theta_1(\pi (z_j-z_k)/L)$, with $\theta_1(z)$ $=$ $\sum_n (-1)^{n-1/2}$ $e^{-\pi (n+1/2)^2}$ $e^{i z (2n+1)}$.  There are $p$ parameters $Z_i$ which represent the location of the center of mass zeros.  In the continuum system there is a symmetry which causes the energy to be independent of how these are chosen.  The space of degenerate states is spanned by $p$ orthogonal wavefunctions.  In most lattice models this symmetry is broken, and the degeneracy is lifted.  Since Eq.~(\ref{wfn}) is made up of lowest Landau wavefunctions, in our model the degeneracy persists.  In Fig.~\ref{4x4} we confirm this degeneracy via an exact diagonalization calculation for 4 bosons on a $4\times4$ lattice with $p=2$ and hard-core repulsion.

Our results gives some insight into recent calculations of Sorensen et al. \cite{sorensen,hafezi}.  They investigated the standard Bose-Hubbard model with nearest neighbor hopping and a uniform magnetic field.  Fixing the filling factor at $\nu=1/2$, they found that when $\phi$ became of order 0.2 the overlap between the exact ground state and the $p=2$ Laughlin state (\ref{wfn}) begins to rapidly decrease.  The characteristic range of hoppings in our model increases with $\phi$ -- and near $\phi=0.2$ the next nearest neighbor matrix element starts to become significant.



\begin{figure}
\includegraphics[width=2.5in]{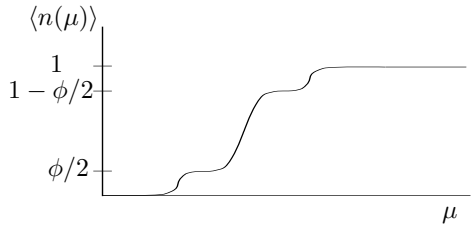}
\caption{Schematic plot of $\avg{n}$ vs. $\mu$, for lattice bosons described by  Eqs.~(\ref{H},\ref{W}), with hard core interactions added.  The steps correspond to incompressible particle/hole fractional quantum Hall states at $\nu = 1/2$. This structure will be visible in the density profile of a trapped gas. Similar structure will be seen with Fermions, but with plateaus at fillings with odd denominators.}\label{steps}
\end{figure}

Since we advocate using cold atoms to investigate this physics, it is important to understand how FQH physics would manifest itself in those systems.  Although most difficult, the most exciting observations would be ones which probe the braiding properties of the excitations \cite{zoller,zhang}.  These states also have definite signatures in Bragg spectroscopy \cite{palmerkleinjaksch}.  The most robust probe, however, is an analog of the vanishing longitudinal resistance seen in solid state systems -- namely the incompressibility of the fractional quantum Hall states \cite{onur,cooper}.  This incompressibility is readily observed in trapped systems, where the chemical potential (and hence the filling factor) varies slowly in space.  As is caricatured in Fig.~\ref{steps}, the equation of state $n(\mu)$ has a series of plateaus corresponding to the filling factor taking on integer fractions.  Within the local density (Thomas-Fermi) approximation, the density profile of the trapped cloud will display these same plateaus \cite{onur,folling,incompressible,oktel}.  The width of these plateaus is set by the gap to single particle excitations in the fractional quantum Hall states.  As shown in figure~\ref{4x4}, in the hard core limit the gap in a $4\times4$ lattice at $\nu=1/2$ and $\phi=1/2$ is 0.566  $t_{nn}$.  This should be compared to the bandwidth $V\approx 4 t_{nn}$.  As $\mu$ goes from 0 to $V$ the density goes from zero to one.  One therefore expects that the $\nu=1/2$ plateau will occupy roughly $1/8$ of the cloud.

{\em Acknowledgements} -- We thank M. Oktel for illuminating discussions. This work was support by NSF grant PHY-0758104 and by the Department of Defense (DoD) through the National Defense Science \& Engineering Graduate Fellowship (NDSEG) Program.




\begin{thebibliography}{99}
\bibitem{trianglefoot} 
 {In this case, we define sublattices by letting all points of the form $\cuof{n+\sqrt{3}i m}$ be in $A$ and $\cuof{(n+1/2)+\sqrt{3}i(m+1/2)}$ be in $B$. We replace $G \of{z}$ by $G_{T} \of{z} = (-1)^{x+y/\sqrt{3}}  \of{z \in A}, (-1) (z \in B),$ and $\pi/2 \to \pi/\sqrt{3}$ in the exponentials.}
\bibitem{entirefoot}{If $f$ diverges faster than $e^{\pi \abs{z}^{2}/2} / z^{2} $, the the sum will \textit{not} be absolutely convergent; we thus can't reorder terms from the Taylor series over $n$ with those from the sum on $z$, invalidating the cancellation order by order in $z$. }
\end{thebibliography}
\end{document}